\documentclass[twocolumn,secnumarabic,amssymb, nobibnotes, aps, prb,superscriptaddress]{revtex4-2}% twocolumn or preprint

\usepackage[pdftex]{graphicx} 
\usepackage{amsmath,nicefrac}
\usepackage{float}
\usepackage[colorlinks=true, linkcolor=blue, citecolor=blue, filecolor=blue, urlcolor=blue]{hyperref}
\usepackage{cleveref}
\usepackage{subfigure}

\usepackage{siunitx}
\begin{document}

\title{An adaptive model for the optical properties of excited gold}

%%%%%%%%%%%%%%%%%%%%%%%%%%%%%%%%%%%%%%%%%%%%%%%%%%%%%%%%%%
\author{P.~D.~Ndione}
\email{ndione@rptu.de}
\affiliation{Department of Physics and OPTIMAS Research Center, RPTU Kaiserslautern-Landau, Erwin-Schr\"odinger-Stra\ss{}e 46, 67663 Kaiserslautern, Germany}
%%%%%%%%%%%%%%%%%%%%%%%%%%%%%%%%%%%%%%%%%%%%%%%%%%%%%%%%%%
\author{S.~T.~Weber}
\affiliation{Department of Physics and OPTIMAS Research Center, RPTU Kaiserslautern-Landau, Erwin-Schr\"odinger-Stra\ss{}e 46, 67663 Kaiserslautern, Germany}
%%%%%%%%%%%%%%%%%%%%%%%%%%%%%%%%%%%%%%%%%%%%%%%%%%%%%%%%%%
\author{D.~O.~Gericke}	
\affiliation{Centre for Fusion, Space and Astrophysics, Department of Physics, University of
	Warwick, Coventry CV4 7AL, United Kingdom}
%%%%%%%%%%%%%%%%%%%%%%%%%%%%%%%%%%%%%%%%%%%%%%%%%%%%%%%%%%
\author{B.~Rethfeld}	
\affiliation{Department of Physics and OPTIMAS Research Center, RPTU Kaiserslautern-Landau, Erwin-Schr\"odinger-Stra\ss{}e 46, 67663 Kaiserslautern, Germany}
%%%%%%%%%%%%%%%%%%%%%%%%%%%%%%%%%%%%%%%%%%%%%%%%%%%%%%%%%%

\date{\today}

%%%%%%%%%%%%%%%%%%%%%%%%%%%%%%%%%%%%%%%%%%%%%%%%%%%%%%%%%%%%%%%%%%%%%%%%%%%%%%%%%%%%%%%%%%%%%%%
\begin{abstract}
We study the temperature-dependent optical properties of gold over a broad energy spectrum covering photon energies below and above the interband threshold. We apply a semi-analytical Drude-Lorentz model with temperature-dependent oscillator parameters. Our approximations are based on the distribution of electrons over the active bands with a density of states provided by density functional theory. This model can be easily adapted to other materials with similar band structures and can also be applied to the case of occupational nonequilibrium. Our calculations show a strong enhancement of the intraband response with increasing electron temperature while the interband component decreases. Moreover, our model compares well with density functional theory-based calculations for the reflectivity of highly excited gold and reproduces many of its key features. Applying our methods to thin films shows a sensitive nonlinear dependence of the reflection and absorption on the electron  temperature. These features are more prominent at small photon energies and can be highlighted with polarized light. Our findings offer valuable insights for modeling ultrafast processes, in particular, the pathways of energy deposition in laser-excited samples.

\end{abstract}
%%%%%%%%%%%%%%%%%%%%%%%%%%%%%%%%%%%%%%%%%%%%%%%%%%%%%%%%%%%%%%%%%%%%%%%%%%%%%%%%%%%%%%%%%%%%%%%

\maketitle

%%%%%%%%%%%%%%%%%%%%%%%%%%%%%%%%%%%%%%%%%%%%%%%%%%%%%%%%%%%%%%%%%%%%%%%%%%%%%%%%%%%%%%%%%%%%%%%
\section{Introduction\label{sec::intro}}
The optical response of materials has been subject to extensive studies over the last decades due to its potential application in various fields, including plasmonics~\cite{hartelt2021, terekhin2022}, material processing~\cite{sugioka2014}, and optoelectronics~\cite{vinnakota2014}. Optical properties are also often used as probes and thus play a crucial role in understanding the fundamental properties of matter. For example, nonequilibrium electron kinetics can be studied in condensed matter using reflectivity and/or transmissivity measurements~\cite{heilpern2018, sun1994, obergfell2020}. Similarly, the combination of optical probing \cite{chen2013, chen2021} and theoretical models~\cite{ndione2022scirep, holst2014} has been used to provide valuable quantitative benchmarks of important properties such as the electron-phonon coupling strength in warm dense matter.  

Laser excitation of matter can strongly modify the energy distribution of its electrons, changing its optical properties, which creates a feedback mechanism to the absorption process. Thus, an accurate description of the transient energy deposition requires an appropriate evaluation of the time-dependent dielectric function. Near equilibrium, it can be provided by  \emph{ab-initio} methods such as density functional theory (DFT) via the Kubo-Greenwood formula \cite{bevillon2018, brouwer2021, silaeva2021, holst2014, zhang2021}. Unfortunately, these methods are not always applicable due to their high computational costs. Moreover, often only one component of the response function is directly calculated~\cite{bevillon2018, holst2014}, while the other component is obtained with the Kramers-Kronig relations, which may lead to increasing inaccuracies. 

For practical reasons, simplified approaches such as the Drude~\cite{drude1900} or the Drude-Lorentz (DL) description are often used. The DL model provides a comprehensive framework for modeling the optical response of various materials as it incorporates interband transitions. It has been widely used~\cite{rakic1998, vial2005, moskovits2002, silaeva2021, rodriguez2017, sehmi2017}, is easy to implement and flexible, \emph{i.e}, it is easy to adapt to different materials, thus, allowing for fast predictions and a quick analysis of experiments. Recently, the  DL model has been fitted to data for gold obtained by DFT calculations \cite{silaeva2021}.

Here, we study the temperature-dependent optical properties of gold across a broad range of probe energies by applying a modified DL model. In contrast to Ref.~\cite{silaeva2021}, we start with a fit to the measured dielectric function at room temperature \cite{johnson1972}. To obtain the optical response at elevated temperatures, we apply temperature-dependent collision frequencies in the Drude part \cite{fourment2014}, {\em i.e.}, the intraband response is fully determined by the temperature-dependent occupation of the conduction band and the changing Drude damping. For the interband response, we introduce an additional temperature dependence of the Lorentz oscillators. In our approach, we connect the amplitudes of the Lorentz oscillators with the density of oscillating electrons at constant oscillator positions. The resonance energies are modified according to the shift of the chemical potential and the broadening of the Fermi edge at elevated temperatures. The remaining freedom for the Lorentz parameters can be fixed by comparison with DFT-based calculations. Finally, the damping of the Lorentz oscillators is described in the same ways as the Drude damping taking into account electron-phonon and electron-electron scattering. In this way, our model allows for a direct transfer of the parameters describing Lorentz oscillators at room temperature to a description applicable to gold with highly excited electrons. 

Our results show a strong enhancement of the intraband response  with increasing electron temperature while the interband component decreases. Most importantly, our model reproduces features in the reflectivity of highly excited gold that have been found with extensive DFT-based calculations~\cite{blumenstein2020}. Applying our methods to thin gold films, we find a strong nonlinear dependence of the optical properties on the electron temperature. We find that the changes are more pronounced at photon energies below the interband threshold and are particularly sensitive to $p$-polarized light, making such a setup ideal for probing.

%%%%%%%%%%%%%%%%%%%%%%%%%%%%%%%%%%%%%%%%%%%%%%%%%%%%%%%%%%%%%%%%%%%%%%%%%%%%%%%%%%%%%%%%%%%%%%%
\section{Theoretical framework}
\label{sec::model}
\begin{figure}[t]
	\centering
	\includegraphics[width=0.48\textwidth]{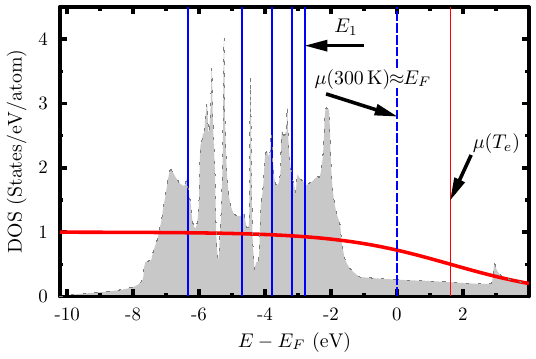}
	\caption{Total density of states of gold. The Fermi energy $E_F$ is plotted as a blue dashed line and sets the zero point of the energy scale. A Fermi distribution for $T_e = \SI{2e4}{\kelvin}$ and the chemical potential at this temperature are plotted with red solid lines.  The resonance energies $E_j$ (Lorentz oscillators) considered in this work are highlighted with blue  solid lines.
	\label{fig::DOS}}
\end{figure}

We consider the excitation of matter by a short pulse of infrared or optical light. For noble metals like gold, such excitations involve only the two upper electron bands, \textit{i.e.}, the $5d$ valence and the $6sp$ conduction electrons. In general, both types of electrons contribute to the optical properties, but the low-frequency behavior is dominated by the $6sp$ electrons~\cite{johnson1972}. The density of states (DOS) can be calculated by density functional theory (DFT). Fig.~\ref{fig::DOS} shows results for gold as obtained by the ELK code~\cite{elk} as well as the occupation of these states for a finite temperature. By projection, partial density of states (PDOS) for $d$- and $sp$-electrons can be obtained as well (see, \emph{e.g.}, Ref.~\cite{ndione2022front}).

The model we present here focuses on time scales when the electron distribution is relaxed and can be described with a temperature. This assumption is often satisfied for excitations with sufficiently high energy, where the kinetic stage establishing a Fermi distribution lasts only a few femtoseconds~\cite{mueller2013, silaeva2018}. Here, the $sp$ and $d$-electrons are assumed to share the same temperature $T_e$ and  chemical potential $\mu$, {\em i.e.}, we have
\begin{equation}
f\!\left(E, \mu,T_e\right)
  = \frac{1}
         {\exp\!\left[\left(E-\mu\right)/k_BT_e\right]
                                                   +1}
\label{eq::Fermi}
\end{equation}
over both bands. Note that this assumption may not hold on time scales when a common electron temperature has been reached but the band occupation has not fully equilibrated~\cite{ndione2022scirep, ndione2019}. The equilibration of the band occupation is, however, much faster than the electron-phonon relaxation, thus there is a stage with a joint Fermi distribution \eqref{eq::Fermi} over both bands where most of the absorbed laser energy is contained in the electronic system. This stage with electron-phonon nonequilibrium is the main focus of interest for our current investigation.

The complex-valued dielectric function is a central and intrinsic quantity to describe how the material responds to an external electromagnetic field, such as the radiation field of a laser pulse. Once known, material properties such as reflection, conductivity, or absorption can be calculated directly~\cite{born1980}. The dielectric function is determined by the electron states specific to the material, {\em i.e.}, the band structure. For metals like gold with a $d$-band character, the total dielectric function is a combination of intra- and interband contributions:  $\varepsilon_{\rm tot} = \varepsilon_{\rm intra} + \varepsilon_{\rm inter}$. While the intraband response is dominated by the $sp$-electrons, the interband response  depends on the occupation within the $d$-band.

%%%%%%%%%%%%%%%%%%%%%%%%%%%%%%%%%%%%%%%%%%%%%%%%%%%%%%%%%%%%%
\subsection{Intraband response}
\label{subsec::intrabandResponse}
In the optical limit with negligible momentum transfer from the photons, the intraband response for the electrons can be modeled with the Drude theory \cite{drude1900}
\begin{equation}
	\label{eq::drude_eq}
	\varepsilon_{\rm intra}\!\left(\omega, T_i, T_e\right) = \varepsilon_{\infty} - \frac{\omega_p^2\!\left(T_e\right)}{\omega^2 + i\omega\nu\!\left(T_i, T_e\right)} \,,
\end{equation}
where $\omega$ is the frequency probed, $\omega_p$ denotes the plasma frequency, and $\nu$ is the collision or damping frequency, which depends on the phonon and electron temperatures $T_i$ and $T_e$, respectively. The background dielectric constant $\varepsilon_{\infty}$ differs from unity in noble metals and is $\varepsilon_{\infty} = 3.104$ for gold~\cite{rodriguez2017}.

The plasma frequency sets the frequency of collective electron excitations and is defined by the density of $sp$-electrons via $\omega_p^2\!\left(T_e\right) = e^2\,n_{sp}\!\left(T_e\right)/\epsilon_0\,m_{sp}^*$ with $e$ being the electron charge, $\epsilon_0$ the vacuum permittivity, and $m_{sp}^*$ the effective mass of the conduction electrons. For the latter, the free-electron mass is a good approximation in gold~\cite{johnson1972}. In our case, the plasma frequency becomes temperature-dependent as thermal excitations increase the $sp$-density
\begin{equation}
	\label{eq::sp_occupation}
	n_{sp}\!\left(T_e\right) =   \int \mathrm{d}E\;f\!\left(E, T_e, \mu\right)\;D_{sp}\!\left(E\right)\,,
\end{equation}
where $D_{sp}\!\left(E\right)$ is the PDOS of the $6sp$-band.

The Drude damping  in Eq.~\eqref{eq::drude_eq} is a key parameter when modeling the optical and transport properties of highly excited metals~\cite{fourment2014, blumenstein2020, ndione2022front}. Its main contributions in gold are electron-phonon and electron-electron scattering rates. Thus, we have $\nu_{\rm tot} = \nu_{ei}\!\left(T_i\right) + \nu_{ee}\!\left(T_e\right)$. At room temperature, it is dominated by the scattering of the $sp$-electrons with the lattice phonons, $\nu_{ei}$, as Pauli-blocking prohibits scattering with $d$-electrons. This rate is set in this work to match the experimental data of Johnson \& Christy~\cite{johnson1972}. For a heated lattice, we assume $\nu_{ei}$ to increase linearly with $T_i$. For the description of $\nu_{ee}$, we follow the model presented in Ref.~\cite{ndione2022front} which takes into account the temperature dependence of the $d$-band hole and $sp$-band electron densities. At elevated $T_e$,  the scattering of $sp$-electrons by $d$-electrons is the dominant damping mechanism and depends strongly on the number of conduction electrons~\cite{fourment2014}, given by Eq.~\eqref{eq::sp_occupation}. Details for the description of the Drude damping are given in Ref.~\cite{ndione2022front}.

%%%%%%%%%%%%%%%%%%%%%%%%%%%%%%%%%%%%%%%%%%%%%%%%%%%%%%%%%%%%%
\subsection{Interband response - room temperature\label{subsec::interbandRoomTe}}
Photons with energies larger than the interband threshold can excite $d$-band electrons into the empty states of the conduction band. Strictly speaking, such interband processes should be treated using the matrix elements that describe the effect of light on the electrons as well as the density of states~\cite{fox2010}. However, the Lorentz model provides a simplified way to describe the interband response and it has been applied frequently to materials at room temperature~\cite{rakic1998, rodriguez2017, navarrete2018}. 

In the Lorentz model, valence electrons are assumed to be bound to the nucleus similar to a spring. The model mimics the motion of electrons around their equilibrium positions with a damped harmonic oscillator driven by an electric field. These oscillators have specific resonance frequency, amplitude, and damping. The inclusion of damping in the Lorentz model implies that the electrons change their state, \textit{e.g.}, via collisions. The damping broadens the absorption line to a finite width~\cite{fox2010}, which also avoids singularities at the resonance frequency. Most materials have multiple characteristic resonances and the total response is a sum over all transitions possible in the optical medium. In the framework of this approach, the interband response can be written as~\cite{rodriguez2017}
\begin{equation}
	\label{eq::Lorentz_neq}
	\varepsilon_{\rm inter}\!\left(\omega\right) = \sum_{j=1}^{\ell}\frac{a_{j,0}}{\omega_{j,0}^2 - \omega^2 - i\omega\Gamma_{j,0}}\,,
\end{equation}
where $a_{j,0}$, $\omega_{j,0}$ and $\Gamma_{j,0}$ are the amplitude, the resonance frequency, and the damping of the Lorentz oscillators, respectively. Here, we choose to apply five Lorentz oscillators. This number is phenomenological and may be optimized to the energy spectrum of interest. An important property of the Lorentz dielectric function is the fact that it fulfills the Kramers-Kronig relations as long as the dampings $\Gamma_{j,0}$ are positive~\cite{moskovits2002, dressel2002}.

For gold at room temperature, we obtain the Lorentz parameters $a_{j,0}$ and $\Gamma_{j,0}$ through a fit to experimental data~\cite{johnson1972}. The resulting parameters are summarized in Tab.~\ref{table}. The five resonance frequencies used here were taken from Ref.~\cite{rodriguez2017}. Fig.~\ref{fig::DOS} shows  the position of the resonance energies $E_j=-\hbar\omega_{j,0}$ (blue vertical lines) in relation to the density of states. Here, and for the values given in Tab.~\ref{table}, we make the Fermi energy $E_F$ the zero point of our energy scale. 

\begin{table}[t]
	\caption{Fit parameters of the Lorentz oscillators at room temperature. $\hbar\Gamma_{j,0}$,  $\hbar\omega_{j,0}$, and $\hbar a_{j,0}$ are given in eV.
	\label{table}}
	\centering
	\begin{tabular}{c|ccccc}
		\hline
		$j$                     & $1$   & \quad$2$   & \quad$3$    & \quad$4$    & $\quad5$     \\
		\hline
		\hline
		$\hbar a_{j,0}$            & 4.577 & \quad4.999 & \quad12.281 & \quad24.999 & \quad 34.999 \\
		$\hbar\Gamma_{j,0}$  & 0.499 & \quad0.699 & \quad0.999  & \quad1.651  & \quad2.190   \\
		$\hbar\omega_{j,0} $ & 2.784 & \quad3.183 & \quad3.799  & \quad4.695  & \quad6.349   \\
		\hline
		\hline
	\end{tabular}
\end{table}

%%%%%%%%%%%%%%%%%%%%%%%%%%%%%%%%%%%%%%%%%%%%%%%%%%%%%%%%%%%%%
\subsection{Interband response - excited matter}
\label{subsec::interbandHighTe}
The application of the Lorentz model to excited matter is not straightforward. As for the intraband response, we have to account for modifications due to the electron and phonon temperatures \cite{ping2006}. In particular, new transitions become possible as an increasing number of free states are created below the Fermi energy when the temperature rises. We propose here to keep the form of the interband response described by Eq.~\eqref{eq::Lorentz_neq}, but introduce temperature-dependent Lorentz parameters. Recently this approach has been applied to electron-phonon nonequilibrium by fitting the Lorentz model to results obtained by DFT at distinct elevated electron temperatures~\cite{silaeva2021}. Our approach aims to introduce a temperature dependence to the strength and damping of the Lorentz oscillators in a smooth way by taking into account the changing occupation numbers in the two optical active bands. The oscillator positions will be kept constant with respect to the density of states. In this way, we provide an easy-to-use description that is smoothly connected to the values at room temperature and might also be applied in nonequilibrium.

We start by considering the amplitudes of the Lorentz oscillators  $a_j$. This parameter can be connected to the number of oscillating $d$-band electrons~\cite{gamboa2015}. Thus, we scale $a_j$ with the electron occupation at the oscillator position in the $d$-band normalized to the value at \SI{300}{\kelvin},
\begin{equation}
	\label{eq::approx_Aj}
	a_j\!\left(T_e\right) = a_{j,0}\;\frac{n_d^{E_j}\!\left(T_e\right)}{n_d^{E_j}\!\left(\SI{300}{\kelvin}\right)}\,.
\end{equation}
Here,  $n_d^{E_j}\!\left(T_e\right)$ is the density of electrons in the $d$-band within an energy interval $\left[E_j - \Delta,  E_j + \Delta\right]$ that contains the energy $E_j$ of the oscillator $j$
\begin{equation}
	\label{eq::n_d_scaling}
	n_d^{E_j}\left(T_e\right) =   \int_{E_j-\Delta}^{E_j+\Delta} \mathrm{d}E\;f\!\left(E, T_e, \mu\right)\;D_{d}\!\left(E\right)\,,
\end{equation}
where $D_d$ is the PDOS of the $d$-band. The parameter $\Delta$ represents a small energy interval. We apply the mean value theorem for integrals to Eq.~\eqref{eq::n_d_scaling} and obtain that the amplitude is proportional to the distribution function at the oscillator position $E_j$
\begin{equation}
	\label{eq::approx_Ai_f}
	a_j\!\left(T_e\right) = a_{j,0}\;\frac{f\!\left(E_j, T_e\right)}{f\!\left(E_j, \SI{300}{\kelvin}\right)}\,.
\end{equation}

Next, we focus on the temperature dependence of the resonance frequencies. When the electron temperature rises, holes are created in the $d$-band below $E_F$ as
more and more $d$-band electrons are thermally excited into the conduction band, which becomes increasingly populated. Moreover, the chemical potential $\mu$ also shifts at elevated temperatures, which reflects the conservation of the total number of particles. The details of the shift depend on the shape of the DOS around $E_F$. In the case of gold, the DOS exhibits high $d$-band peaks below $E_F$ whereas much fewer states exist above the Fermi energy (see Fig.~\ref{fig::DOS}). Thus, the chemical potential for gold rises with increasing electron temperature~\cite{blumenstein2020}. The chemical potential at a temperature of $T_e = \SI{2e4}{\kelvin}$ and the corresponding Fermi distribution are plotted in Fig.~\ref{fig::DOS} (red lines).

This shift of the chemical potential renders interband transitions more difficult as the $d$-electrons need to overcome a broader energy gap. However, the broadening of the Fermi edge facilitates the transitions from $d$- into $sp$-states below the chemical potential as not all of these are fully occupied anymore, {\em i.e.}, we see a softening of the Pauli blocking with rising electron temperature. To account for both effects, the resonance frequencies are modified according to
\begin{equation}
	\label{eq::change_interbandEnergy}
	\hbar\omega_j\!\left(T_e\right) = \hbar\omega_{j,0} + \Delta\chi\!\left(T_e\right) \,,
\end{equation}
where $\Delta\chi\!\left(T_e\right) = \chi\!\left(T_e\right) - \chi\!\left(\SI{300}{\kelvin}\right)$ is the temperature-dependent shift from the fixed basis points of the Lorentz oscillators to the lowest available state for the conduction band. The function $\chi$ is set to be 
\begin{equation}
	\label{eq::chiFunction}
	\chi\!\left(T_e\right) = \mu\!\left(T_e\right) - c\,k_B T_e\,.
\end{equation}
The first term accounts for the shifting chemical potential and the second term for the temperature broadening of the Fermi edge. The constant $c$ sets the strength of the second effect. It can be fixed by comparison to experimental data or first-principles simulations.  

The last quantities to consider are the damping parameters. We describe their temperature dependence similarly to the Drude damping, \textit{i.e.}, including electron-phonon and electron-electron scatterings
\begin{equation}
	\label{eq::lorentz_dampings}
	\Gamma_j\!\left(T_i, T_e\right) = 
	\Gamma^j_{ei}\!\left(T_i\right)
	+ \Gamma_{ee}\!\left(T_e\right) \,.
\end{equation}
At room temperature, the electron-phonon damping $\Gamma_{ei}^j$ dominates and their value is identical to $\Gamma_{j,0}$ presented in Tab.~\ref{table}. For elevated temperatures, one can assume a linear increase with the lattice temperature. However, we consider cases where the probe time is fast and the lattice remains cold, {\em i.e.}, the damping due to scattering with phonons remains unchanged here. In contrast, the damping due to electron-electron scattering $\Gamma_{ee}$ is strongly varying with electron temperature via changes in the occupation of the $d$- and $sp$-bands. Here, we propose to use the same form as for the intraband part, since the damping of the interband transitions relies on the same scattering process~\cite{fourment2014, ndione2022front}. 

To summarize our model, we propose a description of the dielectric function employing the Drude-Lorentz form with temperature-dependent oscillator frequencies, amplitudes, and damping terms, thus, describing intraband and interband contributions. For the Drude part, that is for the intraband contributions, this extension has been shown to work well and to agree with experimental data~\cite{fourment2014, ndione2022scirep}. The Lorentz part, describing transitions between bands, will be tested against DFT-based data in the following section. Despite its simplicity, we will demonstrate that the model is able to reproduce many features of the optical properties of excited gold with very low computational effort.

%%%%%%%%%%%%%%%%%%%%%%%%%%%%%%%%%%%%%%%%%%%%%%%%%%%%%%%%%%%%%%%%%%%%%%%%%%%%%%%%%%%%%%%%%%%%%%%
\section{Results and discussion\label{sec::results}}
We present the results of our analysis of the dielectric function, and derived optical properties, of gold under electron-phonon nonequilibrium. We also compare our results for the reflectivity of gold with calculations based on DFT.  Applying our methods to thin films reveals more detailed insights into the optical properties and points toward potential applications in various fields.

%----------------------------------------%
\subsection{Temperature-dependent dielectric function\label{subsec::dielectricFunction}}
\begin{figure}[t]        
    \centering    
    \begin{minipage}[b]{.5\columnwidth}
       \includegraphics[width=\textwidth]{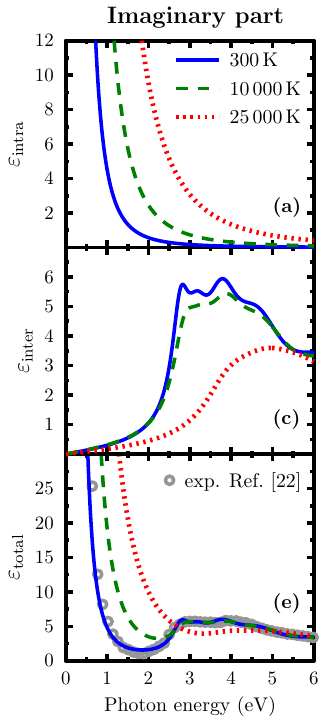} 
       \subfigure{\label{fig::result::1a}}{}
	    \subfigure{\label{fig::result::1b}}{}
	    \subfigure{\label{fig::result::1c}}{}       
    \end{minipage}
    \hfill
    \hspace*{-.2cm}
    \begin{minipage}[b]{.5\columnwidth}
    	\centering    
        \includegraphics[width=\textwidth]{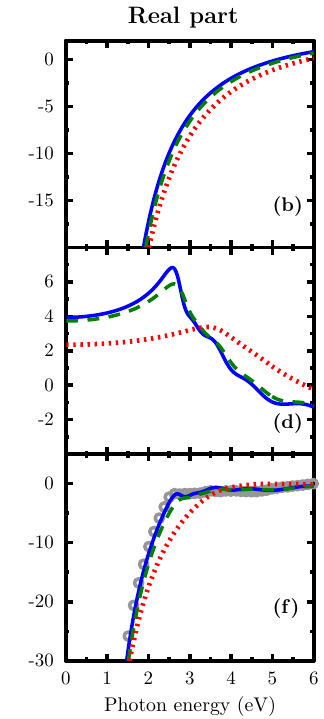}
        \subfigure{\label{fig::result::1d}}{}
		\subfigure{\label{fig::result::1e}}{}
		\subfigure{\label{fig::result::1f}}{}       
    \end{minipage}
    \caption{Broadband dielectric function of gold for various electron temperatures. Imaginary part (left panel) and real part (right panel). \textbf{(a)} and \textbf{(b)} intraband Drude response, \textbf{(c)} and \textbf{(d)} interband Lorentz part, and \textbf{(e)} and \textbf{(f)} total response. The experimental data at room temperature are taken from Ref.~\cite{johnson1972}.\label{fig::result::1}}
\end{figure}

We examine the dielectric function of gold across a broad spectral range to cover absorption of and probing by photons of different energy. Fig.~\ref{fig::result::1} shows the dielectric function in dependence on the photon energy for various electron temperatures while keeping the phonons at \SI{300}{\kelvin}. The individual intra- and interband contributions are shown in Figs.~\ref{fig::result::1a} to \ref{fig::result::1d}. The intraband part is calculated with Eq.~\eqref{eq::drude_eq} and the interband response is determined by Eq.~\eqref{eq::Lorentz_neq}. The total dielectric function, \textit{i.e.} the sum of both contributions, is shown in Figs.~\ref{fig::result::1e} and \ref{fig::result::1f}.

At room temperature and photon energies in the infra\-red regime, the imaginary part of the dielectric function is dominated by intraband contributions, \textit{i.e.}, the Drude response of free carriers as interband transitions are very unlikely, compare Figs.~\ref{fig::result::1a} and \ref{fig::result::1c}. However, this is not true for the real part that has positive and non-negligible interband contributions at low photon energy, as depicted in Fig.~\ref{fig::result::1d} and in qualitative agreement with the findings of Ref.~\cite{ng2016}. That is, the Drude model may not be sufficient to describe the real part of the dielectric function of gold even for the small photon energies.   

With the onset of interband transitions, {\em i.e.,} for photon energies around \SI{1.9}{\eV}, the imaginary part of the inter\-band (Lorentz) part increases significantly as shown in Fig.~\ref{fig::result::1c}. The real part, plotted in Fig.~\ref{fig::result::1d}, shows a maximum at a photon energy equal to the first resonance energy $\hbar\omega_{1,0}$, before decreasing at higher photon energies. In contrast, the imaginary part of the intraband (Drude) contribution decreases monotonically, while its real part increases, see Figs.~\ref{fig::result::1a} and \ref{fig::result::1b}.

The total dielectric response function, Figs.~\ref{fig::result::1e} and \ref{fig::result::1f}, shows a Drude-like behavior for energies below the onset of interband transitions and is relatively flat above $\sim$\SI{3}{\electronvolt}. We take the experimental data of Ref.~\cite{johnson1972} (gray circles) as a reference at \SI{300}{\kelvin} and match these data reasonably well with the fit parameters in Table~\ref{table}.

At elevated electron temperatures, we observe a strong enhancement of intraband contributions in the imaginary part over the entire spectrum , plotted in Fig.~\ref{fig::result::1a} for \SI{10000}{\kelvin} and \SI{25000}{\kelvin}. We attribute this enhancement to a strong rise of both the Drude damping and the plasma frequency \cite{ndione2022front}. In contrast, the temperature effects on the real part of the Drude response are small.

The interband (Lorentz) response also varies with the electron temperatures. It is calculated using Eqs.~(\ref{eq::Lorentz_neq}) to (\ref{eq::lorentz_dampings}) and by setting $c=1/2$ in Eq.~\eqref{eq::chiFunction}, we include the temperature broadening of the Fermi edge in the $sp$-band. The imaginary part becomes smaller for increasing electron temperatures, as can be seen in Fig.~\ref{fig::result::1c}. This behavior is expected: With rising $T_e$ the chemical potential shifts to higher energies, increasing the resonance frequencies and thus decreasing the probability of transitions.  The real part of the interband response decreases at low energies as well. Moreover, the main resonance peak shifts to higher energies (see Fig.~\ref{fig::result::1d}). Due to the softer decrease of the distribution function at the chemical potential, the transition peaks are more and more washed out at elevated $T_e$, leading to smoother interband curves, where the changes in the individual contributions are stronger at photon energies below the interband transition threshold. Figures \ref{fig::result::1e} and \ref{fig::result::1f} show that the total dielectric function is considerably modified at these smaller photon energies, whereas the total response changes just slightly above the threshold. Such behavior is in qualitative agreement with previous observations for copper obtained by the Kubo-Greenwood formalism \cite{bevillon2018}.

\begin{figure}[t]
	\centering
	\includegraphics[width=0.5\textwidth]{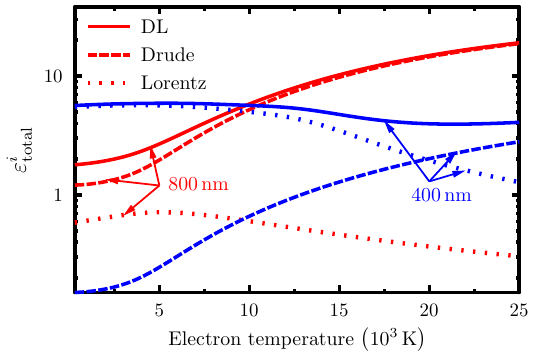}
	\caption{Imaginary part of the dielectric function of Au in dependence of the electron temperature at laser wavelengths of \SI{800}{\nano\meter} (red) and \SI{400}{\nano\meter} (blue). Intraband parts (dashed lines), interband contributions (dots), and total (solid lines). Note the logarithmic  scale on the vertical axis.\label{fig::result::2}}
\end{figure}
\begin{figure*}
	\centering
	\includegraphics[width=\textwidth]{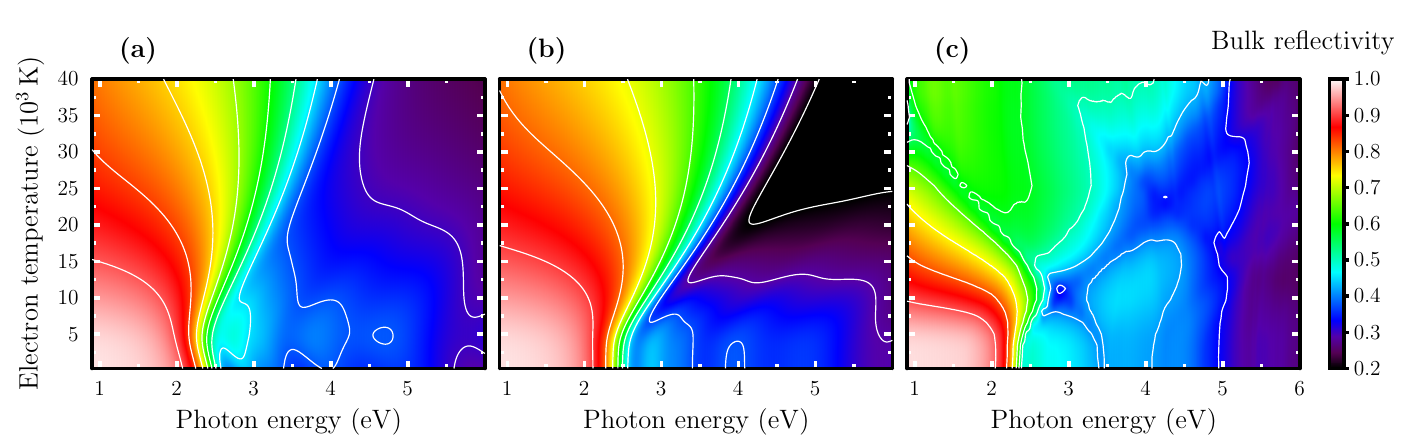}
	\subfigure{\label{fig::result::3a}}{}
	\subfigure{\label{fig::result::3b}}{}
	\subfigure{\label{fig::result::3c}}{}
	\caption{Surface reflectivity of bulk gold versus electron temperature and photon energy.  Panel \textbf{(a)} uses $c=1/2$ in Eq.~\eqref{eq::chiFunction}, \textit{i.e.}, including a temperature broadening of the Fermi edge for the lowest available state. In Panel \textbf{(b)} we consider $c=0$, \textit{i.e.}, the lowest available state in the $sp$-band is determined only by  $\mu\!\left(T_e\right)$. The last panel \textbf{(c)} shows DFT-based calculations from Ref.~\cite{blumenstein2020}. \label{fig::result::3}}
\end{figure*}

Figure~\ref{fig::result::2} shows the important imaginary part of the dielectric function in dependence on the electron temperature $T_e$ for two fixed laser wavelengths: \SI{800}{\nano\meter} (\SI{1.55}{\electronvolt}) and \SI{400}{\nano\meter} (\SI{3.1}{\electronvolt}). At room temperature, the Drude part dominates the total dielectric function $\varepsilon_{\rm total}^{i}$ for the long wavelength, \emph{i.e.} low photon energy. The Lorentz contribution is very small (but finite) in that case. Driven by the Drude contribution, $\varepsilon_{\rm total}^{i}$ increases nonlinearly for increasing electron temperatures. The behavior for the short wavelength of \SI{400}{\nano\meter} is quite different. For low electron temperatures, the Lorentz part dominates and the value of $\varepsilon_{\rm total}^{i}$ is relatively constant, before it starts a minor decline toward a minimum around \SI{20000}{\kelvin}, where the Drude and Lorentz curves intersect. Above this temperature, $\varepsilon_{\rm total}^{i}$ increases again, which is not well visible due to the logarithmic scale of Fig.~\ref{fig::result::2}. This rise is attributed to a strong enhancement of the Drude part.

To understand this behavior, it is essential to assess individual contributions. According to Eq.~\eqref{eq::drude_eq}, $\varepsilon_{\rm intra}^{i}$ is proportional to $\nu$ and $\omega_p^2$. Thus, its trend is fully defined by changes in the collision frequency $\nu$ and the plasma frequency $\omega_p$. At sufficiently high $T_e$, a significant number of $d$-electrons are thermally excited into free states of the $sp$-band and create holes below $E_F$. These holes allow for efficient electron-electron scattering and, thus, considerably increase the Drude damping $\nu$~\cite{fourment2014, ndione2022front}. Combined with an increase in the plasma frequency, the intraband response rises. The interband response, on the other hand, shows an overall decrease with increasing temperature. This results from a lower probability of electron transitions as the chemical potential and, thus, the interband excitation threshold rises with temperature. Here, we note that the resonance frequencies are the most critical parameters in the Lorentz approach.

\subsection{Broadband reflectivity of bulk gold\label{subsec::Map}}
The reflectivity of materials is a measurable quantity and can serve as a benchmark for models. It can be directly calculated from the dielectric function discussed above.  Under normal incidence, the surface reflectivity is determined as
\begin{equation}
	\label{eq::bulkReflectivity}
	R\!\left(\omega, T_i, T_e\right) = \left\lvert\frac{\tilde{n}\!\left(\omega, T_i, T_e\right)-1}{\tilde{n}\!\left(\omega, T_i, T_e\right)+1} \right\rvert^2\,,
\end{equation}
where $\tilde{n}$ is the complex refractive index and is related to the dielectric function via $\tilde{n}=\sqrt{\varepsilon}$.

For the following results, we use the Drude-Lorentz form of the dielectric function developed in Sec.~\ref{sec::model} and, as before, keep the phonons at room temperature. Fig.~\ref{fig::result::3} displays the reflectivity of bulk gold as a function of electron temperature and photon energy. For  Fig.~\ref{fig::result::3a}, we set $c=1/2$ in Eq.~\eqref{eq::chiFunction}, whereas for the results in Fig.~\ref{fig::result::3b} the value $c=0$ is used. This means that the temperature broadening of the Fermi edge is applied in the left panel, whereas it is neglected in the middle panel, where only the shift of  the chemical potential is applied. Accordingly, transitions of $d$-electrons are easier in the former case as additional free states are available around the Fermi edge. We compare both results with data from DFT-based calculations~\cite{blumenstein2020} plotted in Fig.~\ref{fig::result::3c}.

\begin{figure*}[t]
     \begin{minipage}{\textwidth}
        \centering
        \includegraphics[width=\textwidth]{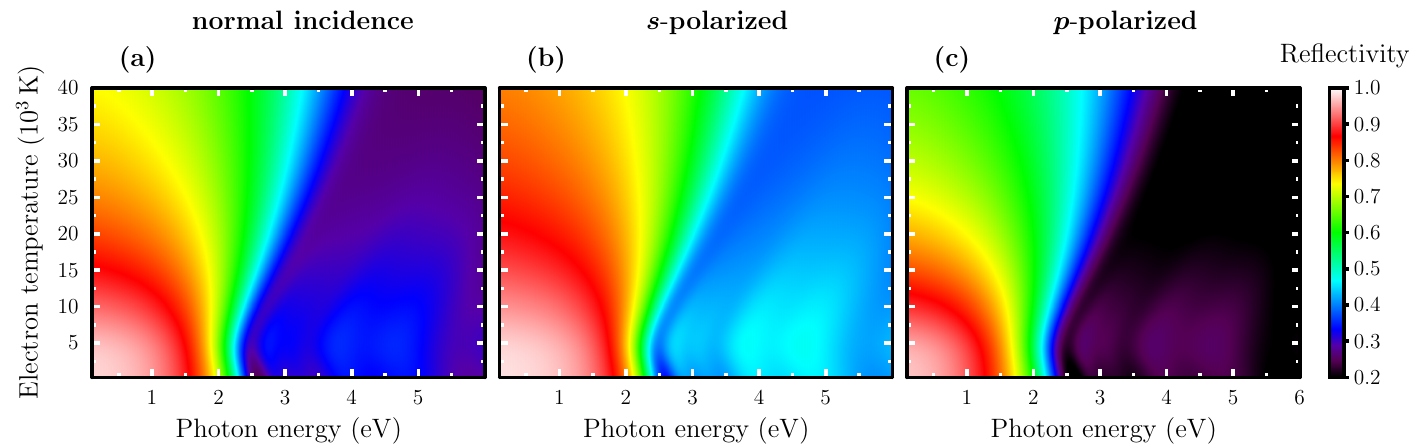}
        \subfigure{\label{fig::result::4a}}{}
	    \subfigure{\label{fig::result::4b}}{}
	    \subfigure{\label{fig::result::4c}}{}
     \end{minipage}\vspace{-1cm}
     \begin{minipage}{\textwidth}
        \centering
        \includegraphics[width=\textwidth]{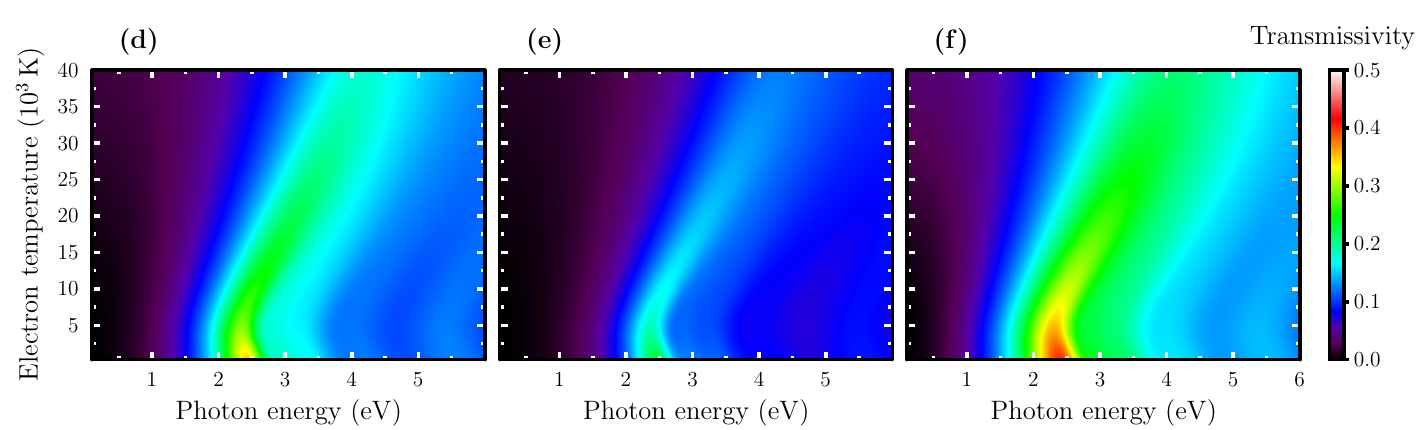}
        \subfigure{\label{fig::result::4d}}{}
	    \subfigure{\label{fig::result::4e}}{}
	    \subfigure{\label{fig::result::4f}}{}
     \end{minipage}\vspace{-1cm}
     \begin{minipage}{\textwidth}
        \centering
         \includegraphics[width=\textwidth]{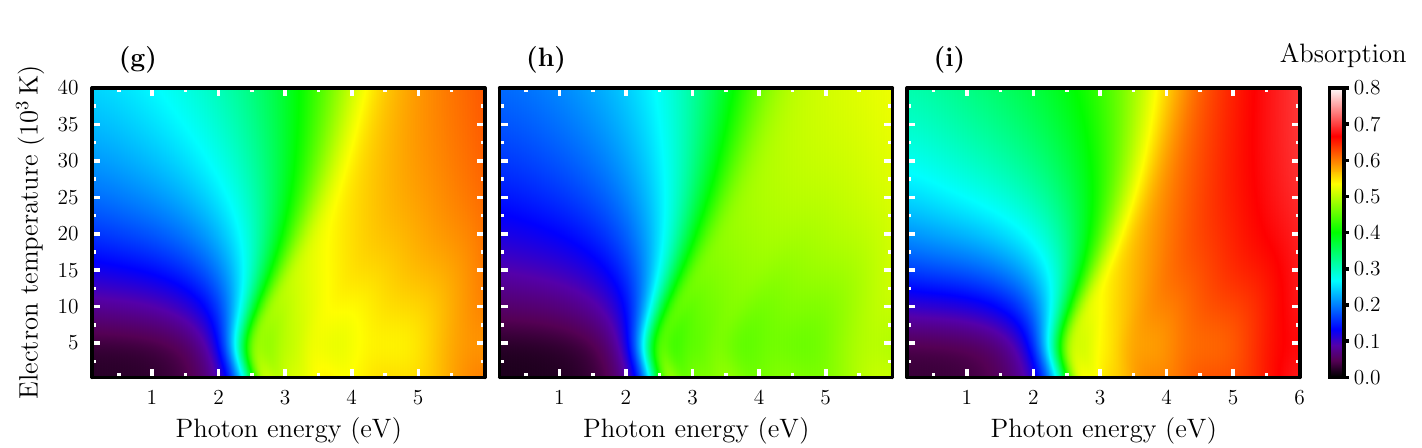}
         \subfigure{\label{fig::result::4g}}{}
	\subfigure{\label{fig::result::4h}}{}
	\subfigure{\label{fig::result::4i}}{}
     \end{minipage}
     \caption{Reflectivity (top row), transmissivity (middle row), and absorption (bottom row) of a \SI{25}{\nano\meter} thin free-standing gold film as a function of electron temperature and photon energy. The results shown in the left vertical column are calculated for light at normal incidence. The panels in the central and right vertical columns are calculated for $s$- and $p$-polarized light, respectively, with an angle of incidence of $\theta_1 = \SI{45}{\degree}$. \label{fig::result::4}}
\end{figure*}

At room temperature, bulk gold is highly reflective below the onset of interband transition energy of $\sim\SI{1.9}{\eV}$.  Above this energy, absorption of photons by $d$-electrons starts to increase strongly and, thus, decrease the reflectivity. With increasing electron temperature, the reflectivity decreases in the infrared regime. Our results depicted in Fig.~\ref{fig::result::3a} show a redshift of the reflectivity edge at temperatures below $\sim\SI{5000}{\kelvin}$. This results from the term $\Delta\chi$ in Eq.~\eqref{eq::change_interbandEnergy} being negative for $c=1/2$ at these low temperatures, \emph{i.e.}, the softening of the Fermi edge dominates the increase in the chemical potential. For higher electron temperatures, the chemical potential increases significantly and shifts the lowest available $sp$-state towards higher energies. Thus, the reflectivity edge experiences a blueshift. In Fig.~\ref{fig::result::3b}, we observe such a blueshift of the reflectivity edge at all temperatures. The blueshift is much more pronounced in Fig.~\ref{fig::result::3b} for the elevated $T_e$ because no broadening of the Fermi edge is considered and, thus, the full energy gap is accounted for. A similar blueshift of the reflectivity edge is also observed in the  DFT-based results plotted  in \ref{fig::result::3c}. Moreover, we see that the edge in the reflectivity is smeared out at elevated $T_e$ both in Figs.~\ref{fig::result::3a} and \ref{fig::result::3b} and also visible in the DFT-based results~\cite{blumenstein2020}. This effect is mainly attributed to the softening of the distribution function. Comparing to the DFT data in panel \ref{fig::result::3c}, we find better agreement of the reflectivity shown in Fig.~\ref{fig::result::3a} for the range of temperatures considered here. Thus, we take $c=1/2$ in Eq.~\eqref{eq::chiFunction} as the best option in future calculations.

Overall, our model shows similar features in the reflectivity as the DFT-based calculations of Ref.~\cite{blumenstein2020} but requires significantly less computational effort. We conclude therefore, that our model provides a reasonably accurate and easily applicable approximation for the behavior of the optical response in highly excited gold including intra- and interband transitions.

\subsection{Application to thin films\label{thinFilms}}
Many experiments applying laser excitation have been performed on thin films. This approach avoids temperature gradients as the foil can be homogeneously heated by ballistic electrons~\cite{chen2021prl,chen2013, hohlfeld2000, chen2012}. To calculate the reflectivity, transmissivity, and absorption of a free-standing thin gold film, we apply multiple reflections theory~\cite{born1980}. We assume a single gold layer with two parallel surfaces and vacuum at both interfaces of the thin film, like, {\em e.g.}, used in Refs.~\cite{chen2021prl, chen2013, ao2006}. We label the vacuum parts with indices $1$ and $3$, respectively, and the gold film is indexed with $2$
leading to~\cite{born1980}
\begin{subequations}
	\label{eq::filmOptics}
	\begin{align}
		R\!\left(\omega, T_i, T_e\right) & = \left\vert\frac{r_{12} + r_{23}e^{i2\beta}}{1 + r_{12}r_{23}e^{i2\beta}}\right\vert^2 \,,
		\label{eq::Ref_films}
		\\
		T\!\left(\omega, T_i, T_e\right) & = \left\vert\frac{t_{12}  t_{23}e^{i\beta}}{1 + r_{12}r_{23}e^{i2\beta}}\right\vert^2 \frac{n_3\cos\!\left(\theta_3\right)}{n_1\cos\!\left(\theta_1\right)}\,,
		\label{eq::trans_films}
	\end{align}
\end{subequations}
where $r_{km}$ and $t_{km}$  are the reflection and transmission coefficients of light propagating from medium $k$ to medium $m$. These coefficients are determined by the Fresnel equations~\cite{born1980}. In our case,  $n_1$ and $n_3$ are the refractive indices of the vacuum and are thus equal. $\theta_1$ is the angle of incidence, and $\theta_3$ is the angle between the transmitted light and the surface normal ($\theta_1=\theta_3$ because media $1$ and $3$ are both vacua). The quantity $\beta = 2\pi h \cos(\theta_2)n_2/\lambda$ denotes the phase change of light resulting from multiple internal reflections. Here, $\lambda$ is the wavelength of the probe pulse, $h$ is the film thickness, and $\theta_2$ is the refracted angle of light within the gold film (medium 2) after entering from vacuum (medium 1). $n_2$ is the complex refractive index of gold calculated with the temperature-dependent dielectric function described above.

Figure~\ref{fig::result::4} shows the reflectivity $R$ (top row), transmissivity $T$ (middle row), and absorption $A = 1-R-T$ (bottom row) of a  free-standing gold film with \SI{25}{\nano\meter thickness as a function} of the probed photon energy and the electron temperature. We present the results for normal incidence in the left columns. For the results shown in the central and right columns, we consider an angle of incidence of $\theta_1 = \SI{45}{\degree}$, where Figs.~\ref{fig::result::4}\textcolor{blue}{(b-e-h)} are for $s$-polarized light and Figs.~\ref{fig::result::4}\textcolor{blue}{(c-f-i)} assume $p$-polarized light.

At \SI{300}{\kelvin}, the thin film reflects less light than the bulk material (compare the top panels to Fig.~\ref{fig::result::3}) for all considered cases. Moreover, the reflectivity edge is lower and less pronounced for the thin film, especially for the $p$-polarized light. As $T_e$ increases, the reflectivity decreases in the infrared regime. The $s$-polarized light is more reflective over the whole spectrum compared to the $p$-polarized light, whose reflectivity shows enhanced features. Similar to bulk, we also observe a blueshift of the reflectivity edge and its broadening at elevated electron temperatures.

The effect of elevated electron temperature on the transmissivity is relatively low, except for photon energies between $\sim\SI{1}{}$ and $\sim\SI{3}{\electronvolt}$. The transmissivity maps all show a peak around $\sim\SI{2.4}{\electronvolt}$, which decreases and shifts towards higher energies with increasing $T_e$. These features appear at normal incidence and for both polarizations considered but are more pronounced for the $p$-polarized light. 

Since the transmissivity is low, the absorption is more or less the inversion of the reflectivity. For small photon energies below the absorption edge, the absorption at room temperature is very low (less than $6\%$), but increases significantly with increasing temperature. In contrast, the absorption is large for higher photon energies, due to the possibility of transferring $d$-electrons into free states of the $sp$-band, and appears to be nearly independent of temperature. Similar to the reflectivity, we also observe a shift of the absorption edge and its broadening.  

Our results for thin films show a similar qualitative behavior of the optical parameters for normal and for oblique incidence, but with large differences in magnitude. Moreover, some of the features observed for $R$, $T$, and $A$ are much more pronounced for $p$-polarized light than for $s$-polarized light. The latter is thus more sensitive and seems to be ideal for probing changes in the optical properties. Figs.~\ref{fig::result::4} also reveal that  photon energies below the interband threshold are better suited to detect changes in material properties than higher photon energies.

\begin{figure}[t]
	\centering
	\includegraphics[width=0.5\textwidth]{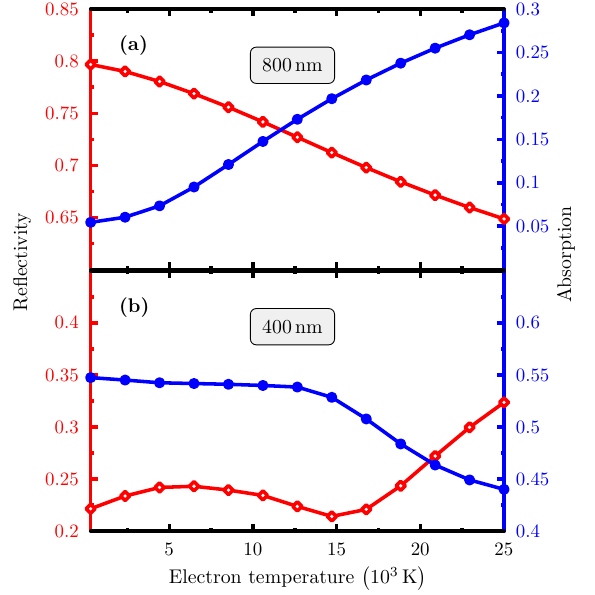}
	\subfigure{\label{fig::result::5a}}{}
	\subfigure{\label{fig::result::5b}}{}
	\caption{Reflectivity and absorption of a \SI{25}{\nano\meter} thin free-standing gold film for $p$-polarized light as a function of the electron temperature. \textbf{(a)} for a wavelength of \SI{800}{\nano\meter} and \textbf{(b)} for \SI{400}{\nano\meter}.\label{fig::result::5}}
\end{figure}

The dependence of the optical properties on electron temperature is difficult to assess quantitatively in Fig.~\ref{fig::result::4}. We thus highlight it for two fixed probe wavelengths, one below and one above the interband threshold. Fig.~\ref{fig::result::5} shows the reflectivity and absorption for $p$-polarized light as a function of the electron temperature for a gold film of \SI{25}{\nano\meter} thickness. We consider probe wavelengths of \SI{800}{\nano\meter} (\SI{1.55}{\electronvolt}) and  \SI{400}{\nano\meter} (\SI{3.1}{\electronvolt}). Depending on wavelength, we observe a different qualitative behavior of reflectivity and absorption. For \SI{800}{\nano\meter}, the reflectivity decreases with increasing $T_e$. This is due to an increase in the number of free states below $E_F$. In contrast, for \SI{400}{\nano\meter}, Fig.~\ref{fig::result::5b} shows a small increase in $R$ when gold is heated. This is followed by a small decrease reaching a minimum at about $\sim\SI{15000}{\kelvin}$ before $R$ experiences a larger increase. The increase in reflection is due to a decrease in the number of free states above the Fermi level with increasing $T_e$.

Equally interesting is the absorption shown in Fig.~\ref{fig::result::5}. For the \SI{800}{\nano\meter} laser, the absorption increases nonlinearly with the electron temperature. At the highest temperature shown, the absorption is more than five times larger than its initial value. That is, the absorption follows a trend similar to that of $\varepsilon_{\rm total}^{i}$  shown in Fig.~\ref{fig::result::2}. At sufficiently high $T_e$, more $d$-electrons, which  serve as a reservoir, are  excited above $E_F$ and are responsible for the observed nonlinearity. For \SI{400}{\nano\meter}, the absorption is nearly constant. The absorption only reduces slightly at very high $T_e$ when many $d$-electrons are thermally excited.

Overall, our calculations show the importance of  considering temperature-dependent optical properties when modeling the optical response of thin films after ultrafast excitation. This behavior is particularly important when estimating the time-dependent absorption during transient energy deposition, especially for thin films. Probing with different polarizations can also help to highlight features arising in heated matter and, thus, can be used for probing the states created. 

%%%%%%%%%%%%%%%%%%%%%%%%%%%%%%%%%%%%%%%%%%%%%%%%%%%%%%%%%%%%%%%%%%%%%%%%%%%%%%%%%%%%%%%%%%%%%%%
\section{Conclusion\label{sec::conclusion}}
In summary, we have studied the optical properties of excited gold for a broad spectrum of probe frequencies and a temperature range from room temperature  up to high electron temperatures. We develop our approach starting with a Drude-Lorentz model fitted to experimental data for room temperature. For excited gold, we generalize the description by introducing temperature-dependent parameters describing interband transition in the Lorentz approach. The main quantities here are resonance frequencies, amplitudes, and scattering frequencies, which depend on electron temperature as well as on the occupation of electron states around the Fermi edge. The temperature dependence of the intraband response, \emph{i.e.}, the Drude part, has been studied previously~\cite{ndione2022front}. Our model shows a good agreement with reflectivity data for highly excited gold obtained by DFT-based calculations~\cite{blumenstein2020}. In particular, it reproduces its key features well but requires much less computational effort. 

When applied to a thin gold film, our method reveals a high sensitivity of the reflectivity and the absorption on the electron temperature. For both quantities, we observe a qualitatively different behavior {for photon energies below and above the threshold for interband transitions.} There is a larger sensitivity to increasing electron temperature for the photon energies below the interband threshold. Moreover, we find pronounced features for $p$-polarized light which, thus, can be considered an ideal tool for probing temperature effects.

The model described here can be easily implemented in numerical simulations and offers valuable insights for modeling ultrafast processes during and shortly after the energy deposition with short laser pulses. The model can be applied to situations with Fermi-distributed electrons as well as to nonequilibrium situations. Although only results for a cold lattice are shown, the effect of an excited phonon system can also be included, \emph{e.g.}, by incorporating additional electron-phonon scattering processes. Extensions of the present work towards materials with a similar band structure, such as copper and silver, are straightforward and  require only the adjustment of a few parameters.

%%%%%%%%%%%%%%%%%%%%%%%%%%%%%%%%%%%%%%%%%%%%%%%%%%%%%%%%%%%%%%%%%%%%%%%%%%%%%%%%%%%%%%%%%%%%%%%

\bibliography{biblio}{}
%%%%%%%%%%%%%%%%%%%%%%%%%%%%%%%%%%%%%%%%%%%%%%%%%%%%%%%%%%%%%%%%%%%%%%%%%%%%%%%%%%%%%%%%%%%%%%%

\end{document}